\documentclass[sigconf, nonacm]{acmart}

\usepackage[ruled, linesnumbered]{algorithm2e} 
\usepackage{amsmath}                           
\usepackage{enumitem}                          
\usepackage{graphicx}                          
\usepackage[framemethod=tikz]{mdframed}        
\usepackage{lscape}                            
\usepackage{multirow}                          
\usepackage{setspace}                          
\usepackage{xcolor}                            
\usepackage{xspace}                            

\SetKwInOut{Input}{Input}
\SetKwInOut{Output}{Output}
\SetKwComment{Comment}{/* }{ */}

\DeclareMathOperator*{\argmax}{arg\,max}

\theoremstyle{definition}
\newtheorem{definition}{Definition}
\newtheorem{mdexample}{\hspace{5pt}Example}

\newtheorem*{mdexample*}{Example}
\newenvironment{example*}{
  \begin{mdframed}[backgroundcolor=teal!12, roundcorner=10pt, linewidth=0pt, innertopmargin=1pt, innerbottommargin=5pt, skipabove=3pt, skipbelow=3pt]
    \begin{mdexample*}}
    {\end{mdexample*}
  \end{mdframed}
}

\newcommand{\nosemic}{\SetEndCharOfAlgoLine{\relax}} 
\newcommand{\dosemic}{\SetEndCharOfAlgoLine{\string;}} 
\newcommand{\pushline}{\Indp} 
\newcommand{\popline}{\Indm\dosemic} 

\newcommand{\sysname}{\mbox{\textsc{OpenForge}}\xspace}
\newcommand{\parwoindent}[1]{\noindent\textbf{#1.}}

\newcommand{\eat}[1]{}

\newcommand\vldbdoi{XX.XX/XXX.XX}

\newcommand\vldbavailabilityurl{URL_TO_YOUR_ARTIFACTS}

\begin{document}
\title{\sysname: Probabilistic Metadata Integration}

\author{Tianji Cong}
\affiliation{%
  \institution{University of Michigan}
}
\email{congtj@umich.edu}

\author{Fatemeh Nargesian}
\affiliation{%
  \institution{University of Rochester}
}
\email{fnargesian@rochester.edu}

\author{Junjie Xing}
\affiliation{%
  \institution{University of Michigan}
}
\email{jjxing@umich.edu}

\author{H. V. Jagadish}
\affiliation{%
  \institution{University of Michigan}
}
\email{jag@umich.edu}



\begin{abstract}
  Modern data stores increasingly rely on metadata for enabling diverse activities such as data cataloging and search. However, metadata curation remains a labor-intensive task, and the broader challenge of metadata maintenance—ensuring its consistency, usefulness, and freshness—has been largely overlooked. In this work, we tackle the problem of resolving relationships among metadata concepts from disparate sources. These relationships are critical for creating clean, consistent, and up-to-date metadata repositories, and a central challenge for metadata integration.

  We propose \sysname, a two-stage prior-posterior framework for metadata integration. In the first stage, \sysname exploits multiple methods including fine-tuned large language models to obtain prior beliefs about concept relationships. In the second stage, \sysname refines these predictions by leveraging Markov Random Field, a probabilistic graphical model. We formalize metadata integration as an optimization problem, where the objective is to identify the relationship assignments that maximize the joint probability of assignments. The MRF formulation allows \sysname to capture prior beliefs while encoding critical relationship properties, such as transitivity, in probabilistic inference. Experiments on real-world datasets demonstrate the effectiveness and efficiency of \sysname. On a use case of matching two metadata vocabularies, \sysname outperforms GPT-4, the second-best method, by 25 F1-score points.

\end{abstract}

\maketitle


\ifdefempty{\vldbavailabilityurl}{}{
\vspace{.3cm}
\begingroup\small\noindent\raggedright\textbf{PVLDB Artifact Availability:}\\
The source code, data, and/or other artifacts have been made available at \url{https://github.com/superctj/openforge}.
\endgroup
}

\section{Introduction}
  \begin{figure}
    \centering
    \includegraphics[width=\columnwidth]{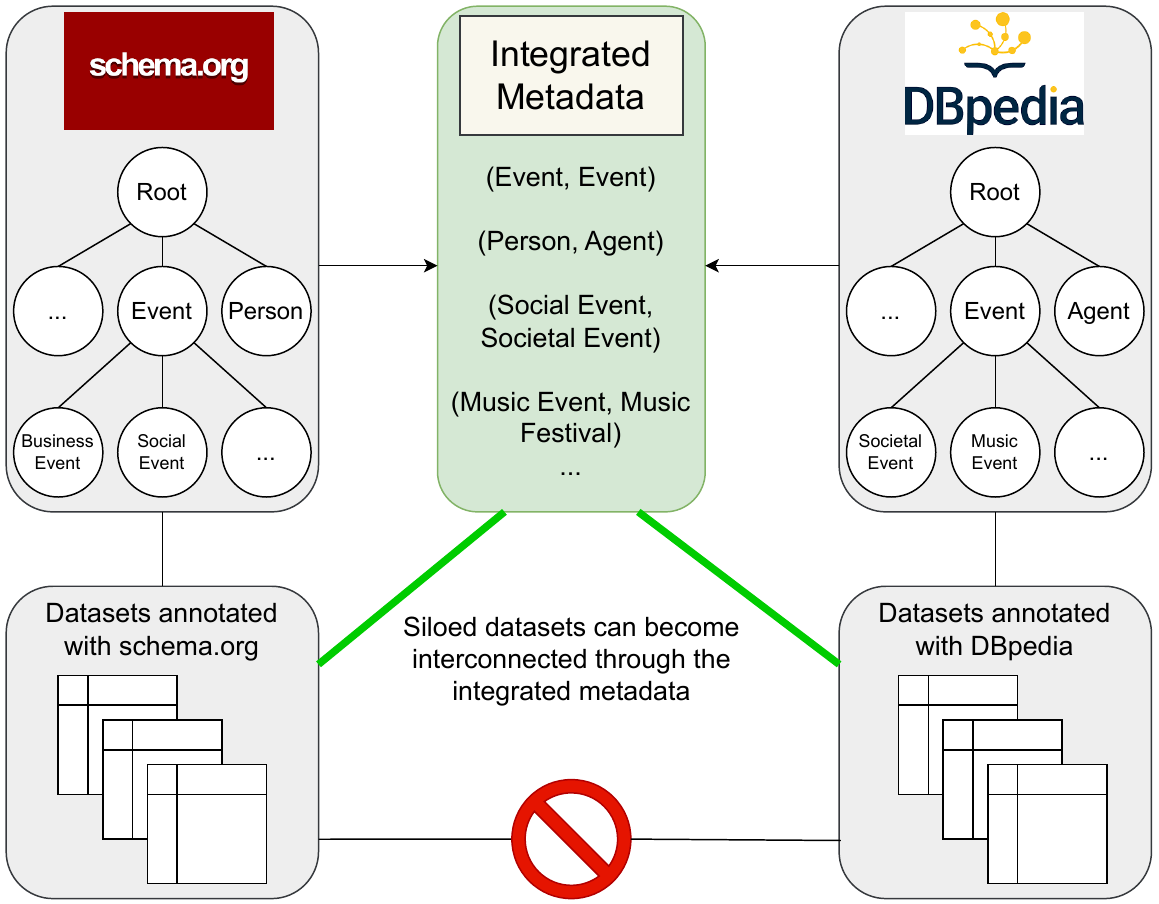}
    \caption{Illustration of metadata integration problem.}
    \label{fig:prob_illustration}
  \end{figure}

  Clean, consistent, and accurate metadata is pivotal in enabling the FAIR Data Principles (findability, accessibility, interoperability, and reusability) in data repositories~\cite{wilkinson2016fair}. For data findability, there has been persistent effort by academia~\cite{NargesianZPM18, BogatuFP020, CasteloRSBCF21, NargesianPBZM23}, industry~\cite{DBLP:conf/sigmod/HalevyKNOPRW16, BrickleyBN19, DBLP:conf/cidr/CongGFJD23}, and open government and scientific data curators~\cite{DataGov, ICPSR}, to improve the performance and efficiency of dataset discovery. While academia has mainly focused on dataset search for data enrichment (joinable and unionable table search), others have worked towards building engines with basic dataset search functionalities. For example, Google dataset search engine~\cite{BrickleyBN19} and \texttt{data.gov}~\cite{DataGov} allow keyword and filter search over dataset descriptions.  
  While there can be some disagreement about the relative importance of such research directions to meet user needs to obtain value from data, there is broad consensus that descriptive, clean, timely, and fresh metadata is critical for data discovery tasks~\cite{DBLP:journals/debu/MillerNZCPA18, DBLP:journals/pvldb/NargesianZMPA19}. 
  Even so, the practice of metadata curation, for dataset publishing as well as dataset indexing, remains a challenge and requires significant human intervention~\cite{DBLP:journals/pacmhci/ThomerAYTPLHY22}. 
  
  \textbf{Metadata Granularity and Vocabulary Mismatch} Google dataset search engine requires data publishers to annotate their datasets with the vocabulary of a known metadata schema, called \texttt{schema.org}. Otherwise, their datasets would not be indexed for search.  
  Web Data Commons~\cite{WebDataCommons}, a long-lasting effort for curating web tables, publishes its benchmarks~\cite{DBLP:conf/semweb/KoriniPB22} annotated with Schema.org types and DBpedia classes~\cite{DBLP:journals/semweb/LehmannIJJKMHMK15}. 
  Open data portals, such as \texttt{CKAN} and \texttt{Socrata}, require each published dataset to be accompanied by a metadata file in \texttt{JSON} format with fields containing metadata information with non-restricted open vocabulary (e.g., \texttt{description}, \texttt{access level}, \texttt{keyword}, \texttt{publisher}, \texttt{theme}, etc.). 
  The metadata schemas and vocabulary domains (\texttt{DBpedia} classes vs. \texttt{schema.org} types vs. open vocabulary) as well as the granularity of provided metadata vary for different datasets, organizations, and repositories. For example, based on our observation, the metadata in \texttt{CKAN} is more coarse-grained and generic than \texttt{Socrata}, making it challenging to find truly relevant datasets to a user's specific query. 
  
  \textbf{Metadata Freshness and Maintenance} A close examination of curatorial work at ICPSR, the world's largest social science data archive~\cite{ICPSR}, reveals that data curation requires curators to use their domain knowledge and make (subjective) judgments to smooth over inconsistencies to achieve standardized metadata~\cite{DBLP:journals/pacmhci/ThomerAYTPLHY22}. 
  For instance, curators draft or revise a dataset's description and create metadata records, such as subject terms and geographic coverage, to support search and retrieval of datasets. Such tasks are commonly backed by metadata standards, like a manually curated vocabulary or thesaurus. However, these standards are not unbreakable rules from the perspective of long-time curators~\cite{DBLP:journals/pacmhci/ThomerAYTPLHY22}. 
  That is, the curated vocabulary and consequently datasets' metadata must be updated to avoid stale metadata and to catch up with the diversity of evolving data. 
  Similarly, curators of data publishing platforms such as \texttt{data.gov} need to integrate various metadata standards, because each state/city independently collects and curates their datasets. 

Maintaining the consistency, usefulness, and freshness of metadata, both within a single repository and across repositories, requires developing techniques to (1) unify and standardize metadata elements, (2) refine the granularity of metadata elements, and (3) perform metadata maintenance holistically at scale. 
Motivated by these challenges, we study 
the problem of integrating metadata from several disparate sources, within a repository or between multiple repositories, to build a consistent and granular view of metadata, by unifying equivalent metadata elements and defining metadata semantic hierarchies. 
Such clean hierarchies can be based on two primary notions to link metadata elements, {\em equivalence} and {\em parent-child} relationships, as seen from the literature on knowledge bases. As illustrated in Figure~\ref{fig:prob_illustration}, given a collection of potentially noisy metadata elements and their accompanying data from different sources, optionally including dataset contents and descriptions, our goal then is to identify all equivalence and parent-child relationships among element pairs. 
Note that while domain discovery~\cite{DBLP:conf/sigmod/ZhangHOPS11, DBLP:journals/pvldb/OtaMFS20} and metadata generation such as column annotation~\cite{DBLP:conf/kdd/HulsebosHBZSKDH19, DBLP:journals/pvldb/ZhangSLHDT20, DBLP:conf/sigmod/SuharaL0ZDCT22}, are relevant tasks, our focus is different: to identify and clean conflicting relationships among a given set of concepts, which could be either raw metadata elements or concepts generated by some algorithms during preprocessing. 

We propose \sysname, a data-driven framework that integrates and resolves inconsistencies among a collection of metadata elements. Conceptually, \sysname's output can be modeled as a graph of metadata elements (concepts) linked with two types of relationships. We call this graph, a {\em relationship assignment graph}. Since the relationship instantiations in this model are unknown apriori, 
we cast the problem of finding an optimal relationship graph to a probabilistic inference problem of relationships. More specifically, we define the problem as a special case of Markov Random Field (MRF), an undirected graphical model. This formulation allows us to (1) plug in various models that leverage metadata elements and accompanying data (if available) for computing priors, (2) resolve inconsistencies by incorporating relationship axioms, such as transitivity, into MRF. 

Our model treats relationships between elements as random variables in MRF and views the assignment of relationships as a probability distribution over random variables. 
Hence, we can define the optimal relationship graph as an assignment of relationships with the maximum joint probability of random variables. MRF by definition decomposes the joint probability to the product of factors (which are non-negative functions over cliques of the graph). 
This lends itself to modeling transitivity with ternary factors (i.e., factors defined over ternary cliques of random variables) to capture the dependency among relationship assignments. 
\sysname allows integrating a wide range of prior models on metadata as well as accompanied data,  
including LLM prompting, LLM fine-tuning, and traditional ML. 
One of the main challenges of applying graphical models to data in the wild is inference scalability~\cite{DBLP:journals/corr/PGMax}. This challenge exacerbates for \sysname when dealing with large datasets with sparse relationships. \sysname overcomes this challenge 
by decomposing the MRF model into smaller graphs, using the transitivity property and metadata similarity, 
and parallelizing inference over these graphs. In summary, we make the following contributions. 
  \begin{itemize}[left=0pt]
    \item We introduce the metadata integration problem as crucial for both unification and integrity maintenance of metadata repositories; we identify equivalence and parent-child relationship determination as central tasks to address this problem, not only to develop relationships but also to get better matches for individual concepts (Section~\ref{subsec:prob_statement}). 
    \item We propose a probabilistic formulation for the problem of resolving equivalence and parent-child relationships among metadata elements (Section~\ref{subsec:prob_formulation}) and cast it as a maximum a posteriori inference problem on Markov Random Field (Section~\ref{subsec:mrf_modeling}).
    \item We propose a data-driven technique, \sysname, that allows incorporating various prior models including LLMs and refining prior predictions through probabilistic inference (Section~\ref{sec:design}).
    \item We mitigate the high time and space complexity of simultaneous probability determination over the entire MRF to enable the scalable inference of relationships for large datasets (Section~\ref{sec:inference}). 
    \item Experiment results on three datasets show that \sysname consistently outperforms task-specific baselines and the latest large language models including GPT-4 by a significant margin. We also demonstrate that MRF inference completes efficiently on our largest dataset and can scale up to graphs with over millions of random variables (Section~\ref{sec:experiments}).
  \end{itemize}


\section{Metadata Integration Problem}\label{sec:prelim}

\subsection{Problem Statement}\label{subsec:prob_statement}

  A metadata repository is the union of the metadata of a collection of datasets. To formalize a metadata repository, we consider metadata elements and their relationships as the building blocks of a metadata repository.  Suppose a set of elements $\mathcal{V} = \{c_{1}, c_{2}, \ldots, c_{n}\}$, where each element represents a unit of metadata information for integration such as column annotations, keywords, entity mentions, or dataset metadata attributes. We call each element in $\mathcal{V}$ a metadata element or concept, interchangeably.  Consider the relationship $R: \mathcal{V}\times\mathcal{V}\rightarrow \{0,1\}$ that can hold between any pair of concepts, e.g., equivalence or parent-child. Our concrete goal is to enrich the metadata of a data repository with a given relationship $R$. More specifically, we want to infer the presence or absence of $R$ between each pair of concepts in $\mathcal{V}$. The concepts in $\mathcal{V}$ are raw metadata units extracted from metadata and dataset files. However, discovering equivalence relationships between pairs of concepts will allow us to unify and standardize these raw concepts. Note that data repositories may contain various types of metadata (e.g., \texttt{keyword}, \texttt{theme}, \texttt{column annotation}). A data curator can choose to integrate specific types or include all metadata, defining $\mathcal{V}$ accordingly. In addition to metadata elements in $\mathcal{V}$, data repositories may also include accompanying data content or descriptions. We refer to this accompanying information, if present, as {\em evidence}, as described in Section~\ref{subsec:prob_formulation}.
  
  In this paper, we focus on two types of relationships: equivalence and parent-child relationships. We note that identifying equivalent elements and parent-child elements is similar to tasks in data matching~\cite{DBLP:journals/pacmmod/TuFTWL0JG23} and taxonomy induction~\cite{DBLP:conf/acl/SnowJN06}, respectively. While data matching solutions typically address schema, entity, and other forms of matching, and taxonomy induction solutions are primarily designed for text and knowledge graphs, our experiments empirically demonstrate why state-of-the-art multi-tasking matching solutions and taxonomy induction approaches fail to effectively merge metadata elements.


\begin{figure}[t!]
  \centering
  \includegraphics[width=\columnwidth]{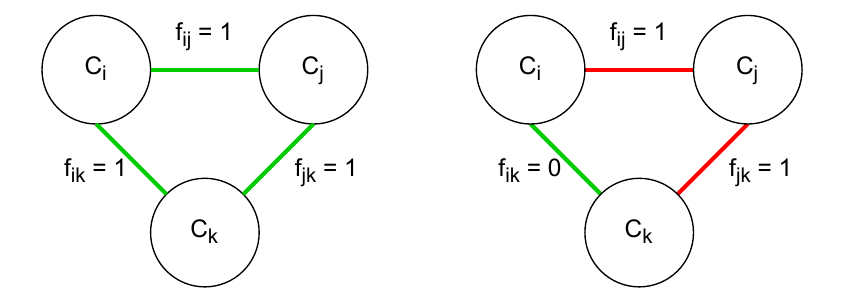}
  \caption{Illustration of the relationship transitivity (left) and inconsistent relationship assignments that violate the transitivity (right). Green edges indicate correct predictions and red edges indicate conflicting predictions.}
  \label{fig:relationship_transitivity}
\end{figure}

\subsection{Probabilistic Formulation}\label{subsec:prob_formulation}
  We first define a relationship assignment graph for relationship $R$, namely $G_R$,  where each node represents a metadata concept in $\mathcal{V}$, and an edge exists between two nodes $c_i$ and $c_j$ if $R$ holds between the corresponding concepts, namely $R(c_i,c_j)=1$. When it is clear from the context, we refer to $G_R$ with $G$. Our goal is to find the most probable relationship assignment graph defined over $\mathcal{V}$.
   
  Let us assume that the presence of the relationship $R$ between any two distinct concepts $c_{i}$ and $c_{j}$ is a random variable $r_{ij}$, following a discrete probability distribution $P(r_{ij})$. This distribution reflects the belief about $c_{i}$ and $c_{j}$ with respect to $R$, where $P(r_{ij} = 0) \,+\, P(r_{ij} = 1) = 1$. Consider a relationship assignment graph $G$. We can define the probability of $G$ as the joint probability of all random variables formed for $\mathcal{V}$ of $G$.

  \begin{definition}[Probability of a Relationship Assignment  Graph]\label{def:prob_relationship_graph}
    \begin{equation*}
      P(G) = P(\, \{ \, r_{ij} \, \vert \, (c_{i}, c_{j}) \in \mathcal{V} \times \mathcal{V}, i < j \, \} \,).
    \end{equation*}
  \end{definition}

  The equivalence relationship is symmetric (i.e., $r_{ij} = r_{ji}$) while the parent-child relationship is not symmetric. For notation convenience, we consider only the random variables for ordered pairs of concepts in Definition~\ref{def:prob_relationship_graph}.

  \textbf{Prior Beliefs.} Although probabilities $P(r_{ij})$ are not known a priori, we assume that for each $r_{ij}$, there exists evidence $e_{ij}$, consisting of features that can be observed (computed) to infer the probability of the concept pair $(c_{i}, c_{j})$ having the relationship $R$. This allows us to compute a prior for $P(r_{ij} \,\vert \, e_{ij})$. We denote the set of observed evidence over all random variables of $G$ as $\mathcal{E}$ and the prior prediction of $r_{ij}$ as $f_{ij}$ where $f_{ij} = \argmax_{r \in \{0, 1\}} P(r_{ij} = r)$. We defer the discussion of obtaining these prior beliefs to Section~\ref{subsec:prior_belief}.
  
  \textbf{Transitivity: Axiom on Relationships.} In addition to the evidence used as priors for relationships, the considered types of relationships in our problem setting, equivalence and parent-child, impose additional structural properties, which can be leveraged for improving relationship predictions. Let us focus on transitivity which introduces dependencies among pairwise relationships. The transitivity condition requires that if a relationship exists between concept pairs $(c_{i}, c_{j})$ and $(c_{j}, c_{k})$, then it must also exist between $(c_{i}, c_{k})$. Formally, considering equivalence or parent-child, for given predictions $f_{ij}$, $f_{ik}$, and $f_{jk}$ evaluated for three concepts $c_{i}$, $c_{j}$, and $c_{k}$, if $f_{ij} = f_{jk} = 1$, transitivity requires that $f_{ik}$ must also be 1. However, if these predictions are made independently, it is possible for $f_{ik}$ to be erroneously predicted as 0, violating the transitivity constraint. The transitivity property of the relationship further implies that certain relationship assignments are inherently inconsistent. Specifically, if we aim to infer $f_{ij}$ and $f_{jk}$ while knowing $f_{ik} = 0$, transitivity dictates that $f_{ij}$ and $f_{jk}$ cannot both be 1, as illustrated in Figure~\ref{fig:relationship_transitivity}. This restriction prevents certain relationship assignments from coexisting. 

  Conflicting predictions can arise if pairwise decisions are made independently and it requires resolution of inconsistencies to maintain transitivity. For example, flipping either $f_{ij}$ or $f_{jk}$ would resolve the conflict by ensuring that the assignments align with the transitivity constraint. Yet, determining which prediction to alter is a non-trivial task, as it may require additional context. To address these challenges, we introduce a probabilistic formulation and a dependency-aware solution below.

  Provided that prior beliefs for each $r_{ij}$,  namely $P(r_{ij} \,\vert \, e_{ij})$, are available and we are given some axioms on the relationships, namely $\mathcal{A}$, the problem of metadata integration can be cast as an optimization problem. The optimization objective is to find the relationship graph assignment $\tilde{G}$ with the maximum probability, conditioned on observed evidence $\mathcal{E}$, while satisfying the constraints implied by the relationship axiom $\mathcal{A}$.

  \begin{definition}[Optimal Relationship Graph]
    Given a relationship $R$, a set of concepts $\mathcal{V}$, and axiom $\mathcal{A}$, the optimal relationship assignment  graph $\tilde{G}$ is the graph with the maximum joint probability of random variables $\{r_{ij}\}_{i<j}$, conditioned on evidence $\mathcal{E}$ and being subject to axiom $\mathcal{A}$.
    \begin{align*}\label{eq:opt_pb}
      \tilde{G} &= \argmax_{G} P(\, G \, \vert \, \mathcal{E}, \mathcal{A} \,) = \argmax_{ \{r_{ij}\}_{i<j} } P(\, \{ r_{ij} \}_{i < j} \, \vert \, \mathcal{E}, \mathcal{A} \,)
    \end{align*}
  \end{definition}

  We consider the common and intuitive transitivity axiom discussed above for our setting. This formulation permits a theoretically optimal solution by exploiting a probabilistic graphical model called Markov Random Field, which also allows for incorporating the relationship transitivity and dependencies during the probabilistic inference. We give a brief introduction to Markov Random Field in Section~\ref{subsec:mrf_modeling}. 

\begin{figure*}[ht!]
  \centering
  \includegraphics[width=\textwidth]{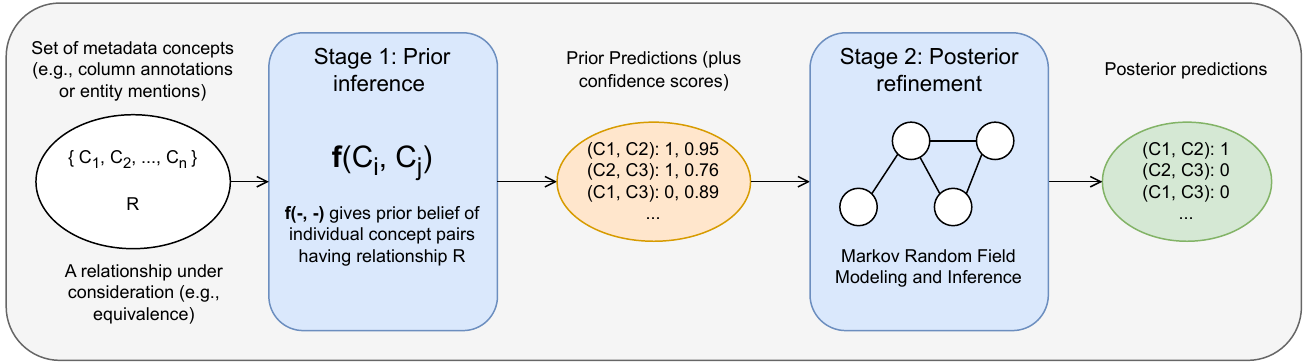}
  \caption{Overview of the proposed two-stage prior-posterior framework for integrating metadata concepts.}
  \label{fig:framework_overview}
\end{figure*}

\section{\sysname Framework}\label{sec:design}
  As illustrated in Figure~\ref{fig:framework_overview}, we introduce \sysname, a prior-posterior framework for integrating metadata concepts. Given a set of concepts and a relationship, the framework operates in two stages. In the first stage, local predictions are generated as prior beliefs for each pair of concepts. Various methods may be used to obtain these prior predictions (Section~\ref{subsec:prior_belief}). In the second stage, we address the relationship transitivity and dependencies between prior predictions through a probabilistic modeling approach using Markov Random Field (Section~\ref{subsec:mrf_modeling}). This graphical model encodes the transitivity constraints and leverages labeled data to learn dependencies in a data-driven manner, effectively capturing the underlying relationship structures and refining the prior predictions.

\subsection{Obtaining Prior Beliefs}
\label{subsec:prior_belief}


  Our probabilistic formulation requires a probability distribution for each random variable. We explore three different methods for obtaining these prior beliefs.

  \subsubsection{Prompting Large Language Models} Recent advances in large language models (LLMs) have shown human-expert performance on a broad array of natural language processing and multi-modality tasks~\cite{DBLP:journals/corr/gpt-4, DBLP:journals/corr/gemini, DBLP:journals/corr/llama3}. These models are pre-trained on a vast amount of curated data and can be applied out-of-the-box to new contexts following a text-to-text paradigm with instruction tuning or prompting~\cite{DBLP:journals/jmlr/RaffelSRLNMZLL20, DBLP:conf/nips/BrownMRSKDNSSAA20}. Motivated by such success, we supply an LLM with a task description and prompt the model for predictions and confidence scores of concept pairs. Besides being straightforward, this method requires no labeled data or only few labeled data for demonstration (i.e., few-shot learning), making it feasible for scenarios where labeled data are scarce or there are not enough data for training/fine-tuning a model.

  \subsubsection{Fine-Tuning Large Language Models} Since LLMs are primarily pre-trained for generation tasks, we also experiment with fine-tuning them for domain-specific classification tasks if training data are available. Due to the large size of LLMs and the hardware constraints, we choose to fine-tune open-source models that have fewer than 10 billions of parameters. In particular, We employ Low-Rank Adaptation (LoRA)~\cite{DBLP:journals/corr/lora}, a parameter-efficient fine-tuning technique that introduces a small set of trainable parameters (known as adapters) while leaving most of the model’s original parameters unchanged. This method enables effective and efficient adaptation of an LLM to a specific relationship classification task with a single GPU. Additionally, LoRA allows for task flexibility: different tasks can share the base LLM, requiring only a switch in adapters rather than maintaining a fully fine-tuned model for each task.

  \subsubsection{Training Machine Learning Models} As compared to the previous methods based on LLMs, we also manually design task-specific features and train a machine learning model, such as Random Forest, to make predictions. This method serves as a baseline to evaluate the LLMs' effectiveness in capturing dataset semantics.

  Further details on the models and adaptations used in our experiments are provided in Section~\ref{subsec:prior_models}. 

\subsection{MRF Modeling of Metadata Integration}\label{subsec:mrf_modeling}

\subsubsection{Markov Random Field Primer}\label{sec:mrfprimer}
  Markov Random Field (MRF), also referred to as Markov Network in the literature, is a specific type of undirected graphical model for representing the joint probability distribution of a set of random variables~\cite{DBLP:books/daglib/Pearl89}. 
  Specifically, an MRF is defined by an undirected graph $M = (\mathcal{N}, \mathcal{L})$ where $\mathcal{N}$ denotes the set of nodes representing random variables and $\mathcal{L}$ denotes the set of edges representing the statistical  dependencies between the random variables.
  
  In MRFs, the joint probability distribution of random variables is factorized as a product of potential functions, each of which is associated with a clique in the graph. The potential function, also known as factor, is a function that assigns a non-negative value to each possible configuration of the random variables in the clique. Here, a configuration of random variables is an assignment of values to random variables. The joint probability of random variables is then proportional to the product of the potential functions over all cliques in the graph, defined as follows. 

  \begin{equation}\label{eq:mrf_joint_prob}
    P(x_{1}, ..., x_{n}) = \frac{1}{Z} \prod_{c \in C} \phi_{c}(x_{c})
  \end{equation} where $x_{i}$ denotes a random variable in 
  $\mathcal{N}$, $C$ denotes the set of cliques in $G$, while $\phi_{c}$ denotes a factor defined over a set of variables $x_{c}$ in a clique $c \in C$, and $Z$ is a normalizing constant. Note that clique set $C$ and potential functions $\phi(c)$ need to be specified per application context.
  
  An MRF can also be represented by a factor graph which explicitly shows the correspondence between random variables and factors. Specifically, a factor graph is a bipartite graph where the nodes are divided into two disjoint sets, one for the random variables and the other for the factors. An edge exists between a factor node and a variable node if the factor depends on that variable. This representation simplifies the computation of probability distributions in MRFs, such as Maximum a Posteriori (MAP) inference, which determines the most likely assignment of values to the variables in the MRF. We introduce our MRF design below and discuss MAP inference in Section~\ref{subsec:mrf_inference}. 


  \begin{figure}[t!]
    \centering
    \includegraphics[width=\columnwidth]{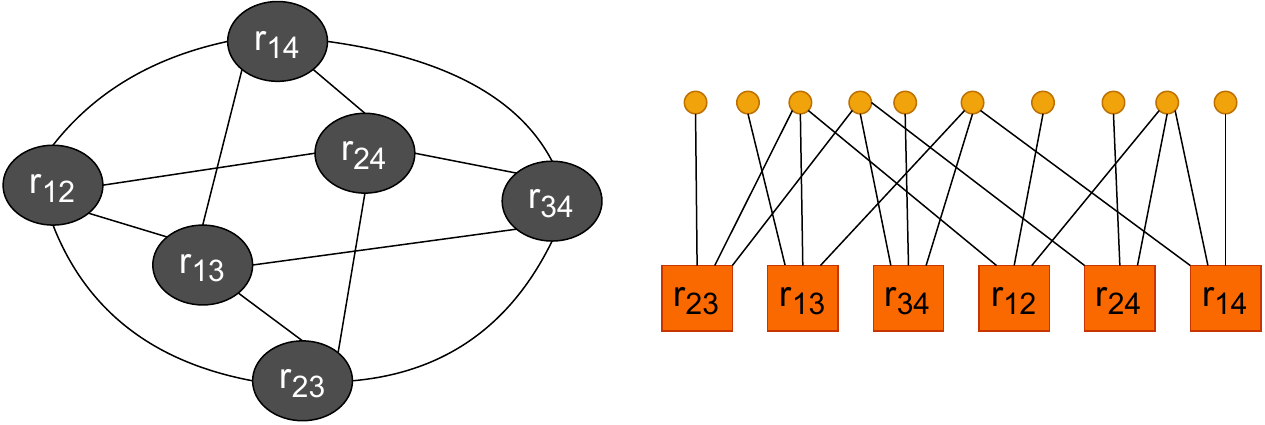}
    \caption{The plot on the left demonstrates an instance of our proposed MRF model containing six nodes/random variables and their dependencies; the plot on the right, known as a factor graph, visualizes the correspondence between factors and random variables in the MRF.}
    \label{fig:mrf_demo}
  \end{figure}

\subsubsection{Our MRF Design}
  MRFs are powerful models for capturing prior beliefs about and the dependencies between random variables. This makes MRFs a natural fit for our problem formulation, which require modeling existing knowledge about concepts and relationships as well as relationship properties such as transitivity.

  Let us first describe how the relationship assignment graph, its probability of Definition~\ref{def:prob_relationship_graph}, as well as relationship transitivity property are modeled as a discrete MRF described in Section~\ref{sec:mrfprimer}. We treat the presence of a specified relationship between a pair of concepts $(c_{i}, c_{j})$, an edge between nodes $c_{i}$ and $c_{j}$ in a relationship graph, as a binary random variable $r_{ij}$. We consider two types of potential functions, or factors, in our problem. The first type is unary cliques $C_{u}$ to capture prior beliefs, each of which contains an individual random variable $r_{ij}$. The second type is ternary cliques $C_{t}$ to capture the relationship transitivity and dependencies, each of which consists of three random variable $(r_{ij}, r_{jk}, r_{ik})$ for $i < j < k$. Note that ternary cliques are the smallest possible cliques on which we can encode the relationship transitivity. 

  The left plot of Figure~\ref{fig:mrf_demo} demonstrates an instance of MRF consisting of six nodes/random variables $\{ r_{ij} \, \vert \, 1 \leq i < j \leq 4\}$ induced by four concepts $\{ c_{1}, c_{2}, c_{3}, c_{4} \}$.  The MRF is densely connected but not fully connected where there is an edge between a pair of nodes if they involve a common concept. For instance, there is an edge between $r_{12}$ and $r_{13}$ because they both involve concept $c_{1}$ whereas there is no edge between $r_{12}$ and $r_{34}$. Under the hood, the edge connectivity is attributed to ternary cliques we consider in the MRF. We do not include all ternary cliques of three random variables in the MRF (e.g., the clique of $r_{12}$, $r_{13}$ and $r_{34}$) but ternary cliques that involve exactly three concepts for the purpose of modeling the relationship transitivity. The right plot of Figure~\ref{fig:mrf_demo} shows the factor graph representation of the MRF that visualizes the correspondence between factors and random variables. This factor graph consists of six random variables and ten factors. Each variable is associated with three factors while each unary factor is associated with one random variable and each ternary factor is associated with a clique of three variables.

  Since an MRF computes the joint probability of the graph based on cliques in the graph, now we can decompose the probability of our relationship assignment graph following Equation~\ref{eq:mrf_joint_prob} as below. 
  \begin{equation}\label{eq:mrf_joint_prob_expanded}
    \begin{split}
      P(G \, \vert \, \mathcal{E}, \mathcal{A})
      &= P(\, \{ r_{ij} \}_{i < j} \, \vert \, \mathcal{E}, \mathcal{A} \,) \\
      &= \frac{1}{Z} \prod_{c \in C} \phi_{c}(\, \{ r_{ij} \}_{i < j, r_{ij} \in c} \, \vert \, \mathcal{E}, \mathcal{A} \,) \\
      &= \frac{1}{Z} \prod_{c \in C_{u}} \phi_{c}(\, r_{ij} \, \vert \, \mathcal{E} \,) \cdot \prod_{c \in C_{t}} \phi_{c}(\, r_{ij}, r_{jk}, r_{ik} \, \vert \, \mathcal{A} \,)
    \end{split}
  \end{equation}

  The potential function for each unary clique $\phi_{c}(\, r_{ij} \, \vert \, \mathcal{E} \,)$ is defined as the conditional probability of the relationship's presence given the observed evidence of the concept pair, allowing us to capture the prior belief of each individual random variable.
  
  While each unary clique has its own potential function, all ternary cliques share a common potential function, which is parameterized to capture relationship transitivity. Specifically, configurations that violate transitivity (i.e., invalid configurations) are assigned a potential value of zero, while each valid configuration is parameterized to represent its likelihood. Higher parameter values indicate a greater probability for the corresponding assignment within the ternary clique. Table~\ref{tab:parameterized_potential_function} gives an example of the parameterized potential function for ternary cliques of binary random variables, where 0 represents no relationship and 1 represents an equivalence relationship. Among the eight possible configurations of three binary variables, three configurations (i.e., [0, 1, 1], [1, 0, 1], and [1, 1, 0]) are invalid and thus set to zero. The remaining configurations are parameterized to determine their potential function values. Although it is technically feasible to assign unique potential functions to each ternary clique, this would result in a model with a linear increase in parameters relative to the number of ternary factors, making the model more complex. To maintain simplicity and prevent over-parametrization, we instead use a shared potential function across all ternary cliques. Hence, Equation~\ref{eq:mrf_joint_prob_expanded} can be written as follows. 
  \begin{equation}
    \begin{split}
      P(G \, \vert \, \mathcal{E}, \mathcal{A})
      &= \frac{1}{Z} \prod_{r_{ij} \in G} P(\, r_{ij} \, \vert \, e_{ij} \,) \cdot \prod_{c \in C_{t}} \phi(\, r_{ij}, r_{jk}, r_{ik} \, \vert \, \mathcal{A})
    \end{split}
  \end{equation}
  where probabilities $P(r_{ij} \,\vert\, e_{ij})$ are given by some prior model and $\phi$ denotes the shared parameterized potential function for ternary cliques, which is independent of prior beliefs.

  The parametrization of the shared potential function for ternary cliques enables us to learn relationship dependencies such as the label distributions in the dataset. In practice, datasets often exhibit skewed class proportions. For example, in an entity matching dataset, there are typically far more non-equivalent pairs of entities than equivalent ones. This imbalance affects the configuration of random variables in ternary cliques. In the parameterized potential function example in Table~\ref{tab:parameterized_potential_function}, the configuration [0, 0, 0] is expected to have a higher potential value than other configurations, reflecting the larger number of concept pairs with no relationship. To learn the parameters in the shared potential function, we treat it as a hyperparameter tuning problem. Rather than manually setting these parameters, we search for well-performing parameters on a labeled validation set using a Bayesian Optimization technique~\cite{DBLP:journals/jmlr/LindauerEFBDBRS22}. We discuss automatic parameter tuning in Section~\ref{subsec:mrf_param_tuning}.

  Provided the learned potential function for ternary cliques, performing Maximum a Posteriori inference over this MRF gives us posterior predictions, an assignment to random variables yielding the maximum joint probability. Algorithm~\ref{alg:mrf_modeling} summarizes our MRF modeling approach. Line 1-2 initialize the MRF with random variables. Lines 3-7 utilize methods, from Section~\ref{subsec:prior_belief}, to estimate the conditional probabilities of random variables, and add unary factors to the MRF. In lines 8-14, we enumerate ternary cliques to create the ternary factors and line 15 executes the inference algorithm.

  We remark that for a set of $n$ concepts, the MRF we propose consists of $\binom{n}{2} = \frac{n(n-1)}{2}$ nodes and $O(n^{3})$ factors (more precisely, $\binom{n}{3}$ factors for ternary cliques plus $\binom{n}{2}$ factors for unary cliques). Section~\ref{subsec:mrf_inference} describes MRF inference and presents an analysis on the time and space complexity of running inference on our proposed MRF model. Section~\ref{subsec:scaling_mrf_modeling} discusses how we scale our MRF modeling for large and sparse datasets.


  \begin{table}[t!]
    \centering
    \caption{Example of the parameterized potential function $\phi$ for ternary cliques of binary random variables $r_{ij}$, $r_{jk}$ and $r_{ik}$ where 0 represents no relationship and 1 represents an equivalence relationship. The potential values $\{ \theta_{i}\}_{i=1}^{5}$ are learnable parameters.}
    \label{tab:parameterized_potential_function}
    \small
    \begin{tabular}{l|llllllll}
    \toprule
    $r_{ij}$ & 0 & 0 & 0 & 0 & 1 & 1 & 1 & 1 \\
    $r_{jk}$ & 0 & 0 & 1 & 1 & 0 & 0 & 1 & 1 \\
    $r_{ik}$ & 0 & 1 & 0 & 1 & 0 & 1 & 0 & 1 \\ \hline
    $\phi(r_{ij}, r_{jk}, r_{ik})$ & $\theta_{1}$ & $\theta_{2}$ & $\theta_{3}$ & 0 & $\theta_{4}$ & 0 & 0 & $\theta_{5}$ \\ \bottomrule
    \end{tabular}%
  \end{table}

\begin{algorithm}[t!]
    \setstretch{1.1}
    \caption{MRF Modeling}\label{alg:mrf_modeling}
    \Input{
        $V$, a set of concepts; 
        $f$, a model that gives prior beliefs; 
        $\phi$, a shared potential function for ternary cliques; 
        $g$, a MRF inference algorithm
    }
    \Output{$G$, a relationship assignment graph}

    \BlankLine
    $mrf \gets init\_mrf(\,)$ \;
    \Comment{Create random variables for each ordered pair of concepts from the initial vocabulary}
    $rvs \gets V$ \;
    \Comment{Add nodes and factors for unary cliques}
    \ForEach{$var \in rvs$}{
        $mrf.add\_node(var), prior\_prob \gets f(var)$ \;
        $unary\_factor \gets init\_factor([var], \, prior\_prob)$ \;
        $mrf.add\_factor(unary\_factor)$ \;
    }
    \Comment{Add edges and factors for ternary cliques}
    $ternary\_cliques \gets generate\_ternary\_cliques(rvs)$ \;
    \ForEach{$(var1, var2, var3) \in ternary\_cliques$}{
        \nosemic $mrf.add\_edges\_from($ \;
          \pushline \dosemic $[(var1, var2), (var2, var3), (var1, var3)])$ \;
        \popline $ternary\_factor \gets init\_factor([var1, var2, var3], \, \phi)$ \;
        $mrf.add\_factor(ternary\_factor)$ \;
    }
    $G \gets g(mrf)$; \Comment{Run MRF inference algorithm}
    \Return{$G$}
  \end{algorithm}

\section{Scalable Metadata Integration}\label{sec:inference}

\subsection{Inference in Markov Random Field}\label{subsec:mrf_inference}
  In our problem, we are interested in finding the most probable relationship graph or equivalently, the most probable assignment to random variables given the observed evidence, i.e., Maximum a Posteriori (MAP) inference. Performing exact MAP inference on complex non-tree structured MRFs is known to be NP-hard and computationally intractable when the graph is large and contains many loops~\cite{DBLP:books/daglib/0023091, StanfordCS228}. For example, a dataset containing 1000 concepts results in half a million of random variables and over 166 millions of factors. We thus resort to approximate inference algorithms, which mainly fall into three categories: linear programming-based inference, sampling-based inference, and variational inference~\cite{StanfordCS228}.

  \subsubsection{Approximate MAP inference.} To find the most efficient existing inference algorithms, we evaluated various implementations from each category of approximate inference algorithms (more details in Section~\ref{subsec:mrf_param_tuning}). We found that Loopy Belief Propagation (LBP)~\cite{DBLP:books/daglib/0023091, DBLP:conf/uai/MurphyWJ99}, a special case of variational inference algorithms, is most efficient and scalable for our MRF model design. Our finding is consistent with the literature where variational inference methods often scale better and are more amenable to optimizations such as parallelization over multiple processors and hardware acceleration using GPUs~\cite{StanfordCS228, DBLP:journals/corr/PGMax}.
  
  LBP is a message-passing algorithm on a loopy graph that iteratively updates the belief of each random variable based on the beliefs of its neighbors in the graph. More precisely, we use the max-product LBP algorithm on factor graphs to perform MAP inference and handle high-order potentials (i.e., potentials of ternary cliques). With respect to the factor graph, LBP computes the message $m_{x \rightarrow w}(x_{i})$ from a variable node $x$ to a factor node $w$ (here factor $w$ contains variable $x$ so there is an edge between the two nodes for message passing) as 
    $m_{x \rightarrow w}(x_{i}) = \prod_{y \in Nb(x) \setminus w} m_{y \rightarrow x}(x_{i})$, 
  where $Nb(x)$ denotes the set of neighbors of node $x$, and $x_{i}$ is a value that $x$ can take. Conversely, the message $m_{w \rightarrow x}(x_{i})$ from factor node $w$ to variable node $x$ is computed as 
    $m_{w \rightarrow x}(x_{i}) = \max_{I, I_{x}=x_{i}} (\phi(I) \prod_{y \in Nb(w) \setminus x} m_{y \rightarrow w}(y_{i}))$, 
  where $I$ is a valid assignment to all variables in factor $w$, tuple ($I, I_{x}=x_{i}$) denotes a valid assignment with $x=x_{i}$, and $\phi$ is the potential function corresponding to $w$. Once a variable $x$ receives messages from all its neighbors in an iteration, the maximal belief of $x$ can be computed as 
    $x^{*} = \argmax_{x_{i}} \prod_{y \in Nb(x)} m_{y \rightarrow x}(x_{i})$. 

  \subsubsection{Time Complexity Analysis.} Having described the MRF approximate inference algorithm we apply for finding the optimal relationship assignment graph, let us describe its complexity. Let $s$ and $d$ be the number of possible states and degree of variable $x$, respectively. Computing a variable-to-factor message $m_{x \rightarrow w}(x_{i})$ requires combining all incoming messages from neighboring factor nodes excluding $w$ for each state of $x$. Then the cost of computing a single variable-to-factor message is $O(s \cdot (d - 1))$. It follows that the time complexity for computing all variable-to-factor messages in one iteration is $O(\sum_{x} d \cdot s \cdot (d - 1)) = O(|V| \cdot s \cdot d^{2})$ where $|V|$ denotes the number of variable nodes in the factor graph.

  Let $d_{w}$ be the degree of factor node $w$. Computing a factor-to-variable message $m_{w \rightarrow x}(x_{i})$ requires maximizing over the values of all variables connected to $w$ excluding $x$. Then, the cost of computing a single factor-to-variable message is $O(s^{d_{w} - 1})$ as there are $s^{d_{w} - 1}$ configurations of connected variables excluding $x$. It follows that the time complexity for computing all factor-to-variable messages in one iteration is $O(\sum_{w} d_{w} \cdot s^{d_{w}})$. Combining the complexities for both types of messages, the time complexity of running max-product LBP for $l$ iterations is
    $O(l \cdot (|V| \cdot s \cdot d^{2} + \sum_{w} d_{w} \cdot s^{d_{w}})).$ 
  
  In our proposed MRF model, there are $\binom{n}{2}$ binary variable nodes with degree $n-1$, $\binom{n}{2}$ unary factors with degree 1 and $\binom{n}{3}$ ternary factors with degree 3 given a set of $n$ concepts. By plugging in these numbers, the time complexity above can be simplified to $O(l \cdot n^{4})$.

  \subsubsection{Space Complexity Analysis} Recall from Figure~\ref{fig:mrf_demo} that factors are modeled as bipartite graphs between variables and factors. Each edge in the factor graph requires two messages---one message from the variable node to the factor node and another from the factor node to the variable node. Storing a message requires space proportional to the number of states of a variable. Then, the space complexity of storing messages is $O(|E| \cdot s)$,  where $|E|$ denotes the number of edges in the factor graph. Additionally, for each variable node, we store a belief over its possible states, which requires $O(s)$  space per variable. For $|V|$ variable nodes, the space required for storing beliefs is $O(|V| \cdot s)$. Thus, the overall space complexity for max-product LBP is
    $O(|E| \cdot s + |V| \cdot s)$
  which translates to $O(n^{3})$ in our proposed MRF model.

  The quartic time complexity and cubic space complexity is forbidding for large datasets with thousands of concepts. Next, we introduce a solution to scale up the MRF inference.

  \begin{figure}[t!]
    \centering
    \includegraphics[width=\columnwidth]{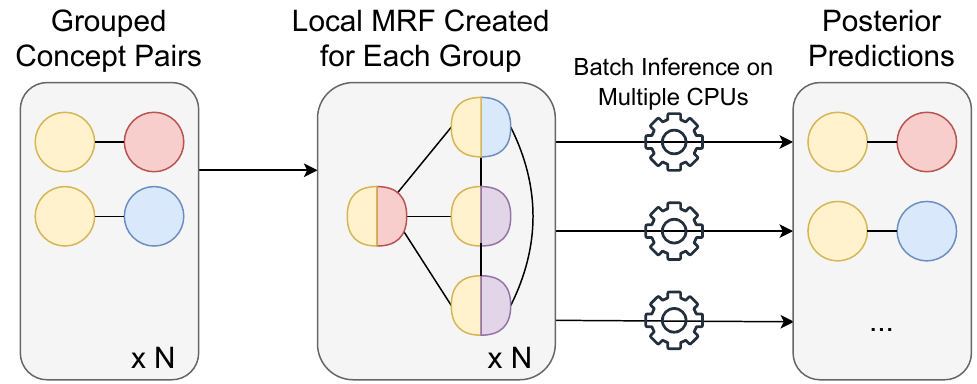}
    \caption{Creating independent MRFs for concept pairs in large datasets with sparse relationships. Ordered pairs are first grouped by the left concept and top-$k$ neighbors (represented by the purple semi-circles) are retrieved for the left concept to construct a local MRF of random variables. Inference over independent MRFs is parallelized on available CPUs and posterior predictions of test pairs are collected.}
    \label{fig:scaling_demo} 
  \end{figure}

\subsection{Scaling MRF Modeling}\label{subsec:scaling_mrf_modeling}
  The high time and space complexity stems from the densely connected graph we construct. For a set of $n$ concepts, each variable node in the resulting factor graph connects to $n-1$ neighbors including one unary factor and $n-2$ ternary factors. This dense connectivity leads to significant computational demands and memory consumption.

  However, we empirically observe that many connections in the dense graph are unnecessary for ensuring relationship transitivity and learning dependencies. Specifically, numerous connections are either redundant or represent weak dependencies that do not significantly impact inference accuracy. We propose to systematically reduce the connections for each node. By reducing the connections from $n-1$ to a constant $k$, where $k << n$, we effectively decrease the number of message-passing operations required by the LBP algorithm, enabling faster execution and significantly lower memory usage. Our intuition is that concepts with more similar embeddings are more likely to share strong relationship dependencies. Therefore, to select these $k$ neighbors for each concept, we employ a text embedding model\footnote{\url{https://huggingface.co/nvidia/NV-Embed-v2}} to represent concepts in embedding space and search for $k$-nearest neighbors based on Cosine similarity. By focusing only on the top-$k$ neighbors, we maintain the essential relationship dependencies while disregarding redundant connections that contribute little to inference quality.

  Reducing the number of connections for each node has the advantage of dividing the MRF into disconnected local graphs. Our solution first groups ordered concept pairs by the left concept and constructs a local MRF for each group. This ensures that the same concept pair does not appear in multiple MRFs and local MRFs are independent of each other. The inference for each local MRF can then be performed independently in constant time. Furthermore, we can parallelize the inference for local MRFs across the available CPUs, which significantly reduce the inference time. Figure~\ref{fig:scaling_demo} provides an illustration of the idea.
  
\section{Implementation}\label{sec:impl}
  We describe implementation details of \sysname in this section.

\subsection{Prior Models}\label{subsec:prior_models}


  \subsubsection{Large Language Models} We employ two open-source LLMs as prior models for prompting and fine-tuning: \texttt{gemma-2-9b-it} from Google~\cite{DBLP:journals/corr/gemma} and \texttt{Qwen2.5-7B-Instruct} from Alibaba~\cite{DBLP:journals/corr/qwen2}. Both models can be accessed from Hugging Face's model hub. We choose these two models as we are able to deploy them locally and they can closely follow our straightforward prompts to generate both predictions and confidence scores. We refer to both models as gemma-2 and qwen2.5 respectively for convenience below.

  Due to the space limitation, we have made our prompts and fine-tuning hyperparameters publicly available in our GitHub repository~\footnote{\url{https://github.com/superctj/openforge}}. Both models can be fine-tuned efficiently under an hour (for 20 epochs). One noteworthy issue we encountered when fine-tuning gemma-2 and qwen2.5 is that both models tend to exhibit excessive confidence in their predictions, with most predictions yielding a probability score above $0.99$. This overconfidence hinders the message-passing process in MRF inference, as only information from the dominant class gets propagated. To mitigate the issue, we incorporate three techniques for calibrating model confidence. Specifically, we use a weighted loss function (with weights inversely proportional to the frequency of the respective classes) when fine-tuning models on imbalanced datasets and enable label smoothing~\cite{DBLP:conf/nips/MullerKH19} in the loss calculation, which assigns a small probability to incorrect classes to reduce overfitting. Additionally, we apply temperature scaling~\cite{DBLP:conf/icml/GuoPSW17} to adjust the sharpness of the output logits during inference, which involves dividing the logits by a temperature value greater than 1 before applying the softmax function, thereby softening the resulting probabilities. 

  \subsubsection{Machine Learning Models} We train a multi-class classification model on a predefined set of features to predict the presence of a relationship for each concept pair. We employ three classifiers including Ridge classifier, Random Forest, and Gradient-Boosted Decision Tree from the scikit-learn library~\cite{DBLP:journals/jmlr/PedregosaVGMTGBPWDVPCBPD11}. These models were chosen for their built-in regularization mechanisms, which help reduce overfitting and improve generalization on class imbalanced datasets. We define features using concept names and associated data values if available. These features include string similarities (Q-gram, Jaccard and edit distance), embedding similarities (fastText~\cite{DBLP:journals/tacl/BojanowskiGJM17}), word count ratio, and character count ratio between concept names, and Jaccard similarities and embedding similarities between column values. 

\subsection{MRF Inference and Parameter Tuning}\label{subsec:mrf_param_tuning}
  \eat{
  \begin{table}[t!]
    \renewcommand{\arraystretch}{1.2}
    \centering
    \caption{Parameters to tune in our problem.}
    \label{tab:param_to_tune}
    \resizebox{\columnwidth}{!}{%
      \begin{tabular}{ll}
        \toprule
        Parameter Source & \multicolumn{1}{c}{Description} \\ \hline
        MRF modeling & \begin{tabular}[c]{@{}l@{}}Up to 2\textasciicircum{}3=64 parameters in the shared potential\\function of ternary cliques.\end{tabular} \\ \hline
        \multirow{3}{*}{\begin{tabular}[c]{@{}l@{}}MRF inference \\ (Loopy Belief Propagation)\end{tabular}} & \begin{tabular}[c]{@{}l@{}}damping: The damping factor to use for message\\updates between one iteration and the next.\end{tabular} \\ \cline{2-2} 
        & \begin{tabular}[c]{@{}l@{}}temperature: A normalizing constant to use when\\computing estimated marginal probabilities.\end{tabular} \\ \cline{2-2} 
        & num\_iters: Number of iterations to run LBP. \\ \bottomrule
      \end{tabular}%
    }
  \end{table}
  }

  We experimented with five implementations of MAP inference over MRFs from three libraries, including the Shafer-Shenoy algorithm~\cite{DBLP:journals/amai/ShaferS90} from \texttt{pyAgrum}~\cite{DBLP:conf/pgm/DucampGW20}, Belief Propagation~\cite{DBLP:books/daglib/Pearl89}, Max-Product Linear Programming (MPLP)~\cite{DBLP:conf/nips/GlobersonJ07}, and Gibbs Sampling~\cite{DBLP:journals/pami/GemanG84} from \texttt{pgmpy}~\cite{ankan2015pgmpy}, and Loopy Belief Propagation (LBP)~\cite{DBLP:conf/uai/MurphyWJ99} from \texttt{PGMax}~\cite{DBLP:journals/corr/PGMax}. 
  We found that LBP from \texttt{PGMax} is overall the most efficient and scalable algorithm for our problem due to their flat array-based implementation of LBP in \texttt{JAX}~\cite{jax2018github}, which uses just-in-time compilation and parallelization to optimize array operations on multi-processors and hardware accelerators such as GPU.

  
  There are two sources of parameters we tune for our problem: the parameters in the potential function of ternary cliques and the parameters of the LBP inference algorithm (e.g., the damping factor used to improve the convergence of LBP). In particular, we tune the parameters using SMAC~\cite{DBLP:journals/jmlr/LindauerEFBDBRS22}, a robust and flexible Bayesian Optimization framework for hyperparameter optimization. SMAC treats the MRF inference algorithm as a black-box function and iteratively searches for a configuration of specified parameters to maximize an objective given by a target function. The target function takes in a configuration of parameters and returns a performance value, e.g., accuracy or F1 score, by running the inference algorithm with the configured parameters. Internally, SMAC employs a surrogate model to approximate the target function and in each optimization iteration proposes a configuration with the maximum expected improvement (which is computed using the surrogate model) for evaluation. The optimization process balances between exploration (i.e., search an unknown region in the parameter space) and exploitation (i.e., search for configurations near the best-so-far configuration), and stops after a fixed number of iterations. SMAC is known to be effective and efficient for AutoML applications~\cite{DBLP:journals/jmlr/LindauerEFBDBRS22} and under active maintenance.
\section{Experiments}\label{sec:experiments}
  We demonstrate the effectiveness, efficiency, and scalability of \sysname in this section. We first describe the experiment setup including datasets, baselines and evaluation metrics in Section~\ref{subsec:exp_setup}. We then present the results of the experiments on the effectiveness of \sysname compared to baselines (Section~\ref{subsec:quality_comparison}), the improvement of \sysname over various prior models (Section~\ref{subsec:prior_models_evaluation}). the efficiency of MRF inference algorithms (Section~\ref{subsec:exp_efficiency}), and the scalability of our MRF modeling (Section~\ref{subsec:exp_scalability}).

\subsection{Experimental  Setup}\label{subsec:exp_setup}

  \begin{table}[t!]
    \renewcommand{\arraystretch}{1.5}
    \centering
    \caption{Summary of dataset statistics. All three datasets are highly imbalanced with positive class proportions under 10\%.}
    \label{tab:dataset_stats}
    \resizebox{\columnwidth}{!}{%
    \begin{tabular}{lccc}
    \toprule
    \textbf{Datasets}                       & \textbf{SOTAB}  & \textbf{Walmart-Amazon} & \textbf{ICPSR}     \\ \midrule
    \multicolumn{1}{l|}{Relationship Type}  & Equivalence     & Equivalence             & Parent-Child       \\
    \multicolumn{1}{l|}{}                   & \multicolumn{3}{c}{Training / Validation / Test}               \\
    \multicolumn{1}{l|}{Number of Concepts} & 46 / - / 46     & 5126 / 2507 / 2484      & 140 / 54 / 54      \\
    \multicolumn{1}{l|}{Number of Pairs}    & 1035 / - / 1035 & 6144 / 2049 / 2049      & 9730 / 1431 / 1431 \\
    \multicolumn{1}{l|}{Prop. of Positive Class} & 1.26\% / - / 1.26\% & 9.38\% / 9.42\% / 9.42\% & 1.63\% / 4.05\% / 3.98\% \\ \bottomrule
    \end{tabular}%
    }
  \end{table}

\subsubsection{Datasets}\label{subsec:exp_datasets}

  We evaluate \sysname on three datasets involving two types of relationships: equivalence relationships (SOTAB dataset for column types matching and Walmart-Amazon dataset for entity matching) and parent-child relationships (ICPSR dataset for taxonomy induction). We have sourced SOTAB and ICPSR datasets particularly for the metadata integration task. 

  \textbf{SOTAB Dataset.} Korini et al.~\cite{DBLP:conf/semweb/KoriniPB22} introduce the Schema.org Table Annotation Benchmark (SOTAB) for understanding the semantics of table elements. We take advantage of the fact that the second version of this  benchmark\footnote{\url{http://webdatacommons.org/structureddata/sotab/v2/}} provides manually verified column type annotation from two vocabularies, i.e., \texttt{schema.org}~\cite{Schemaorg} and DBPedia~\cite{DBLP:journals/semweb/LehmannIJJKMHMK15}.  Hence, we consider an initial collection of concepts by merging  the set of schema.org concepts and DBPedia concepts. If a table column is annotated with concepts from both vocabularies, we consider these concepts as a pair of equivalent concepts. The many-to-many correspondences  from columns to concepts can challenge the priors. This allows us to test how the proposed MRF formulation can identify conflicting relationships. Due to the limited size of this initial vocabulary, we split it to training and test datasets of equal size. 

  \textbf{Walmart-Amazon Dataset.} This dataset has been used to evaluate entity matching algorithms~\cite{DBLP:conf/sigmod/MudgalLRDPKDAR18, DBLP:journals/pacmmod/TuFTWL0JG23}. It provides a challenging setting (e.g., noisy attributes and missing data) for matching product entities between two major e-commerce platforms. We consider this dataset primarily because it is used by one of our baselines, Unicorn, the SOTA for data matching tasks. Moreover, it contains a large number of equivalence relationships. We use the same training, validation, and test splits (at ratio of 0.6:0.2:0.2) as prior work. 

  \textbf{ICPSR Dataset.} ICPSR, the institute maintaining the world's largest social science data archive~\cite{ICPSR}, makes their controlled vocabularies publicly available. We use a subset of the Subject Thesaurus\footnote{\url{https://www.icpsr.umich.edu/web/ICPSR/thesaurus/10001}} from ICPSR to create a dataset of parent-child relationships. The concepts in this dataset are social science subject terms from various disciplines including political science, economics, education, etc. We first manually identify concepts with transitive parent-child relationships such as \texttt{credentials $\rightarrow$ academic degrees $\rightarrow$ doctoral degrees}, then divide the collection of  transitive concept tuples into training, validation, and test datasets at ratio of 0.6:0.2:0.2. This dataset has no column data associated with each concept as ICPSR does not use subject terms to annotate table elements. The prior beliefs of this dataset can be only learned from concept names, which is a more challenging setting.

  Table~\ref{tab:dataset_stats} summarizes the dataset statistics. It is notable that all datasets have highly imbalanced class distributions. The proportion of the positive class in test splits ranges from 1.26\% to 9.42\%, which poses a significant challenge for any solution.

\subsubsection{Baselines.} For each task defined by a relationship, we compare our approach with the previous state-of-the-art (SOTA) methods as well as the latest LLMs, GPT-3.5 and GPT-4, from OpenAI. 

  \textbf{Unicorn.} Unicorn~\cite{DBLP:journals/pacmmod/TuFTWL0JG23} is a unified, multi-tasking model designed to support diverse data matching tasks in data integration (e.g., entity matching and schema matching). It leverages multi-task learning and a Mixture-of-Experts architecture to enable knowledge sharing across tasks and datasets. Additionally, Unicorn claims to support zero-shot prediction for new tasks without requiring labeled data. In our experiments, we compare \sysname with Unicorn on datasets involving equivalence relationships.

  \textbf{Chain-of-Layer.} Zeng and Bai et al. recently propose Chain-of-Layer~\cite{DBLP:conf/cikm/chain-of-layer} for automatic taxonomy induction from a given set of entities. It heavily relies on prompting a LLM to construct the taxonomy from top to bottom in an iterative manner and leverages another smaller language model to reduce hallucinations of the LLM such as removing non-existent entities and parent-child relationships generated by the LLM. We compare \sysname with Chain-of-Layer on datasets involving parent-child relationships. 

  \textbf{GPT-3.5/GPT-4.} The GPT series of models have demonstrated that scaling up language models in size and data significantly improves task-agnostic performance with task and few-shot demonstrations~\cite{DBLP:conf/nips/BrownMRSKDNSSAA20, DBLP:journals/corr/gpt-4}. In our experiments, we prompt GPT-3.5 and GPT-4 to predict the presence of a specified relationship between concept pairs using task descriptions and few-shot learning. We use the OpenAI API to interact with the models identified as \texttt{gpt-3.5-turbo-\\0125} and \texttt{gpt-4-turbo-2024-04-09}. 

  Due to class imbalance in the datasets, we evaluate the quality of all approaches using F1 score, precision, and recall. We report the best results of each approach. For example, LLMs can obtain higher scores without few-shot learning in some cases. Additionally, we report the runtime of our MRF modeling approach in seconds for efficiency and scalability analysis. 

\subsubsection{Hardware} 

Experiments involving GPU-accelerated MRF inference, fine-tuning and inference of open-source LLMs, and runtime measurement are done on a node with a NVIDIA A40 40GB GPU, 16 cores of Intel Xeon Gold 6226R processors and 256 GB of RAM on a shared computing cluster. The rest of experiments are conducted on a local server with 16 cores of Intel Xeon Bronze 3106 processors and 256 GB of RAM. 

\subsection{Quality Comparison}\label{subsec:quality_comparison}
  \begin{figure*}[t!]
    \centering
    \includegraphics[width=\textwidth]{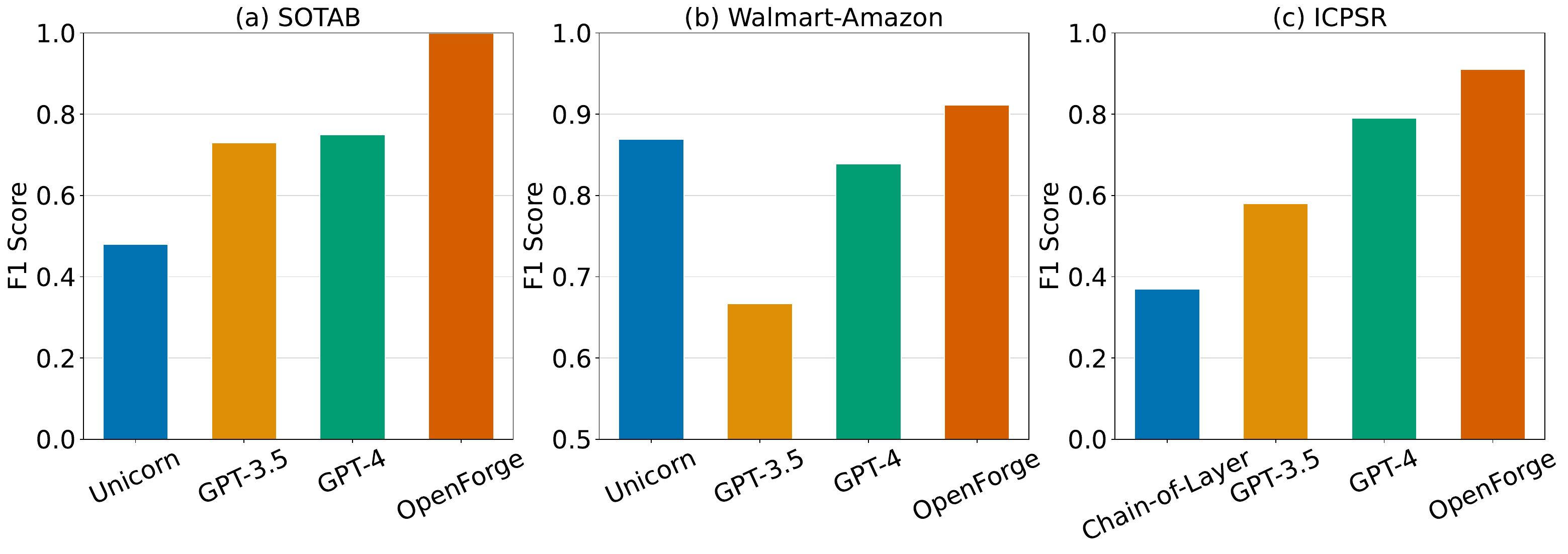}
    \caption{Comparison of F1 score across methods on the SOTAB, Walmart-Amazon, and ICPSR datasets.}
    \label{fig:all_f1}
  \end{figure*}

  \textbf{RQ1: How does \sysname compare with baselines for various metadata integration tasks?}

  \sloppy{We evaluate \sysname alongside baselines on three datasets involving two types of relationships: equivalence relationships (SOTAB dataset for column types matching and Walmart-Amazon dataset for entity matching) and parent-child relationships (ICPSR dataset for taxonomy induction). For the equivalence relationship datasets, the baselines include Unicorn, GPT-3.5, and GPT-4, while for the parent-child relationship dataset, the baselines include Chain-of-Layer, GPT-3.5, and GPT-4. As shown in Figure~\ref{fig:all_f1}, \sysname consistently outperforms the baselines with substantial improvements in F1 score across all datasets. These results highlight \sysname's flexibility and robustness in capturing different relationship types, outperforming both specialized and general-purpose models. We attribute the superior performance to three key factors:}
  \begin{enumerate}[left=0pt]
    \item Joint Probability Modeling: MRFs effectively model the joint probability distribution over relationship variables, enabling more accurate predictions.
    \item Preservation of Transitivity: By incorporating potential functions for ternary cliques, \sysname ensures that invalid assignments violating relationship transitivity are avoided.
    \item Dependency Learning: Fine-tuning hyperparameters of MRF modeling and inference on a validation dataset with a class distribution similar to the test set allows \sysname to capture relationship dependencies more effectively.
  \end{enumerate}

  Table~\ref{tab:quality_comparison} reports detailed quality metrics for all methods and datasets. Among the baselines, GPT-4 consistently outperforms GPT-3.5 across all datasets. This is expected as GPT-4 is significantly larger than GPT-3.5 in model size and have been shown to perform better than GPT-3.5 on various NLP tasks~\cite{DBLP:journals/corr/gpt-4}. Below, we discuss some highlighted results on each dataset.

  \subsubsection{SOTAB dataset} \sysname achieves a perfect F1 score of 1.0 even though the prior beliefs used as inputs to MRF modeling are far from perfect. We report a detailed comparison of prior models and \sysname (prior beliefs + MRF-based modeling) in Section~\ref{subsec:prior_models}. These results provide strong evidence for the advantage of modeling relationship transitivity and dependencies using MRFs. Among the baselines, GPT-4 is the second-most effective method but trails \sysname by 25 F1 points. Unicorn, the previous SOTA for data matching tasks, performs the worst on this dataset. The less ideal performance suggests that Unicorn, despite leveraging multi-task learning, finds it challenging to generalize to the metadata integration tasks and datasets. This is most likely due to two reasons: (1) the SOTAB dataset is not seen in pretraining of Unicorn and (2) equivalent pairs in Unicorn datasets often share common parts whereas column values associated with equivalent types in the SOTAB dataset are heterogeneous and may have no overlap.

  \subsubsection{Walmart-Amazon dataset} \sysname achieves an F1 score of 0.91, outperforming Unicorn, the second-best method, by 4 F1 points. The performance gap is smaller on this dataset as Unicorn was pre-trained on multiple entity matching datasets including the Walmart-Amazon dataset and remains a competitive baseline. Additionally, the sparsity characteristic of this dataset (i.e., many entities appear only once in the test set and the predictions of many entity pairs are independent of others) reduce the advantage of explicitly modeling relationship transitivity and dependencies. Among other baselines, GPT-4 lags behind \sysname by 7 F1 points behind while GPT-3.5 trails by 24 F1 points.

  \subsubsection{ICPSR dataset} Unlike the previous two datasets involving equivalence relationships, the ICPSR dataset focuses on parent-child relationships. In this setting, \sysname outperforms GPT-4 by 12 F1 points and GPT-3.5 by 33 F1 points. The specialized approach, Chain-of-Layer, performs the worst, reflecting its limitations when applied to concepts in the ICPSR dataset that come from diverse disciplines in the social science domain. Chain-of-Layer was originally designed for taxonomy induction in narrow domains, such as the parent-child relationships between countries and provinces/states, and struggles to generalize to broader contexts.


  {\begin{table}[t!]
    \renewcommand{\arraystretch}{1.5}
    \centering
    \caption{Comparison of F1 score, precision, and recall for all methods and datasets. \sysname achieves the best F1 scores across three datasets, which are highlighted in bold.}
    \label{tab:quality_comparison}
    \resizebox{\columnwidth}{!}{%
    \begin{tabular}{lllll}
    \toprule
    \textbf{Dataset} & \multicolumn{4}{c}{\textbf{F1 / Precision / Recall}} \\
                   & \multicolumn{1}{c}{Unicorn}        & \multicolumn{1}{c}{GPT-3.5} & \multicolumn{1}{c}{GPT-4} & \multicolumn{1}{c}{OpenForge} \\ \midrule
    SOTAB          & 0.48 / 0.62 / 0.38                 & 0.73 / 0.60 / 0.92          & 0.75 / 0.63 / 0.92        & \textbf{1.00} / 1.00 / 1.00               \\
    Walmart-Amazon & 0.87 / 0.90 / 0.85                 & 0.67 / 0.57 / 0.80          & 0.84 / 0.76 / 0.93        & \textbf{0.91} / 0.96 / 0.87            \\ \midrule
                   & \multicolumn{1}{c}{Chain-of-Layer} & \multicolumn{1}{c}{GPT-3.5} & \multicolumn{1}{c}{GPT-4} & \multicolumn{1}{c}{OpenForge} \\
    ICPSR          & 0.37 / 0.53 / 0.28                 & 0.58 / 0.54 / 0.63          & 0.79 / 0.70 / 0.91        & \textbf{0.91} / 0.98 / 0.84            \\ \bottomrule
    \end{tabular}%
    }
  \end{table}}

  \begin{figure*}
    \centering
    \includegraphics[width=\textwidth]{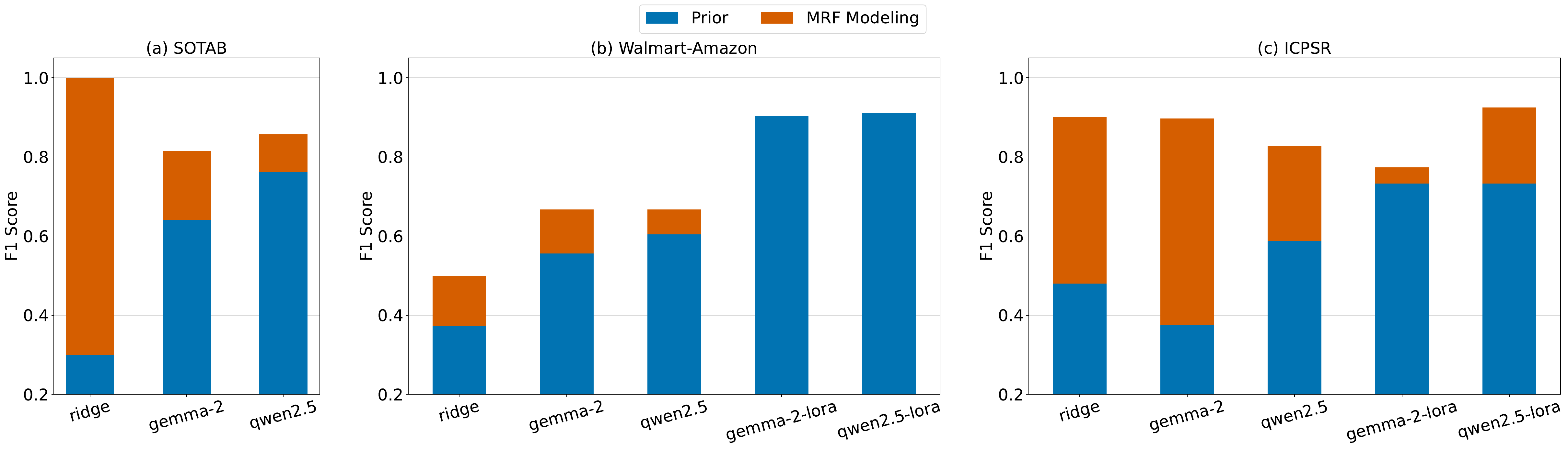}
    \caption{Comparison of F1 scores between prior models and \sysname (prior models + MRF modeling) on (a) SOTAB, (b) Walmart-Amazon and (c) ICPSR datasets. We omit gemma-2-lora and qwen2.5-lora on SOTAB due to insufficient training data.}
    \label{fig:prior_impact}
  \end{figure*}

\subsection{Prior Models}\label{subsec:prior_models_evaluation}
  \textbf{RQ2: Can MRF modeling improve on various prior beliefs and by how much?}

  Figure~\ref{fig:prior_impact} illustrates the impact of prior models on the result quality of \sysname for the SOTAB, Walmart-Amazon, and ICPSR datasets. To save space, we present results for up to five prior models: the Ridge classifier (the top-performing ML model among three tested), gemma-2, qwen2.5, and their LoRA fine-tuned classifier variants, i.e., gemma-2-lora and qwen2.5-lora. On all datasets, our MRF modeling significantly improves the results quality over the prior models, with gains ranging from 6 to 70 F1 points. These results demonstrate the effectiveness of MRF modeling in enhancing prior predictions and correcting mispredictions that violate relationship transitivity.

  We observe that LoRA fine-tuned LLMs outperform vanilla LLMs with few-shot learning by a large margin. For instance, the LoRA fine-tuned gemma-2 model achieves an F1 improvement of 52 points compared to the vanilla gemma-2 model, while the corresponding improvement on the Walmart-Amazon dataset is 45 F1 points. This highlights the potential of fine-tuning vanilla LLMs to obtain strong performance for classification tasks. Note that on the Walmart-Amazon dataset, MRF modeling does not provide additional improvements for the gemma-2-lora and qwen2.5-lora models. This is attributed to two factors: (1) the dataset’s sparsity, with many entity pairs in the test set being independent, thereby limiting the occurrence of relationship transitivity violations; and (2) fine-tuned LLMs as strong priors, with manual checks confirming that none of their mispredictions violate relationship transitivity. Nevertheless, applying MRF modeling does not degrade the quality of the results, preserving the performance of these strong prior models.

  \textbf{Guide for Practitioners.} When sufficient training data (e.g., thousands of examples) are available, fine-tuning LLMs using parameter efficient fine-tuning techniques for domain-specific tasks yields very strong results and is highly recommended. In cases where training data are lacking, prompting a LLM to obtain prior beliefs offers a viable alternative. 


  \begin{figure*}[ht!]
    \begin{minipage}[t]{\columnwidth}
      \centering
      \includegraphics[width=\columnwidth]{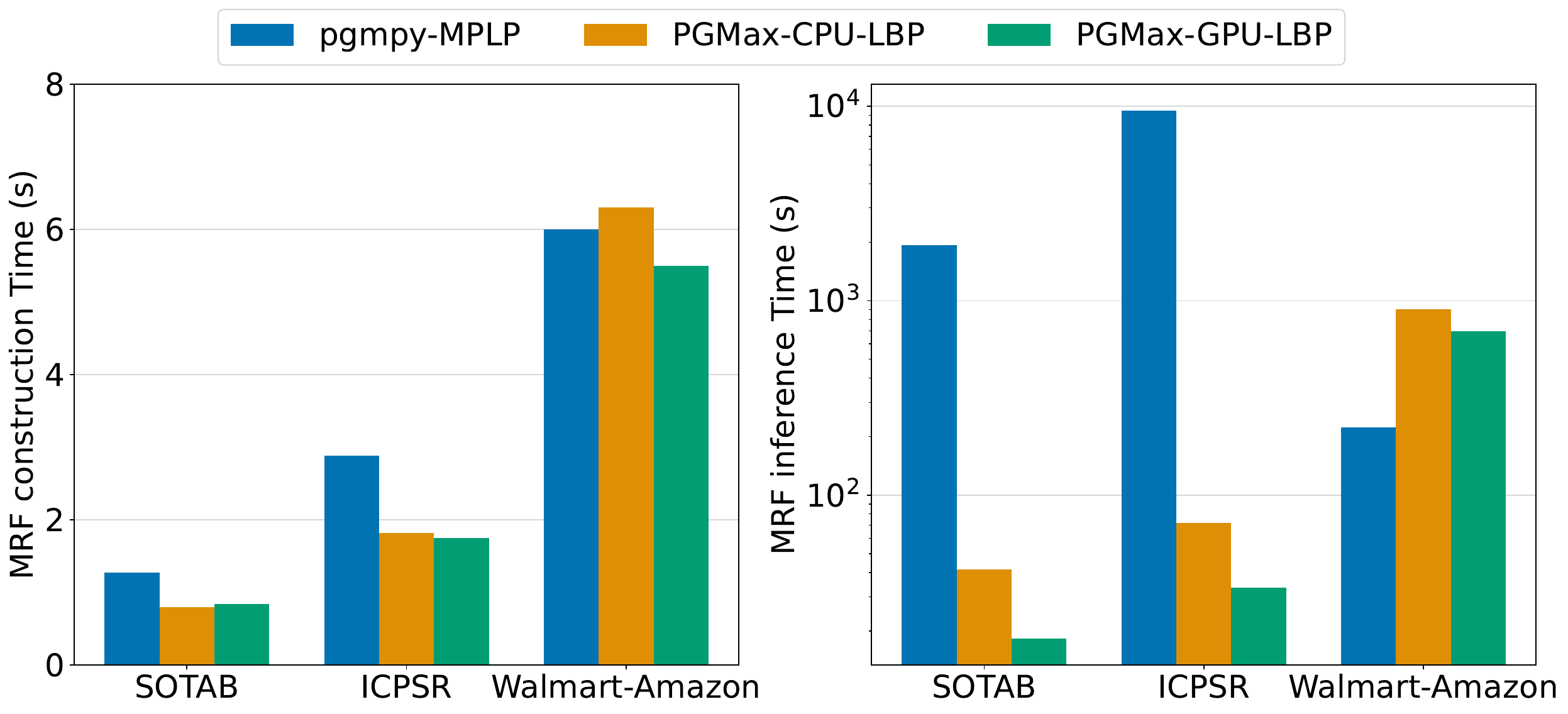}
      \caption{MRF construction and inference time of three inference algorithms across datasets using a single CPU or GPU.}
      \label{fig:efficiency}
    \end{minipage} \hfill
    \begin{minipage}[t]{\columnwidth}
      \centering
      \includegraphics[width=0.95\columnwidth]{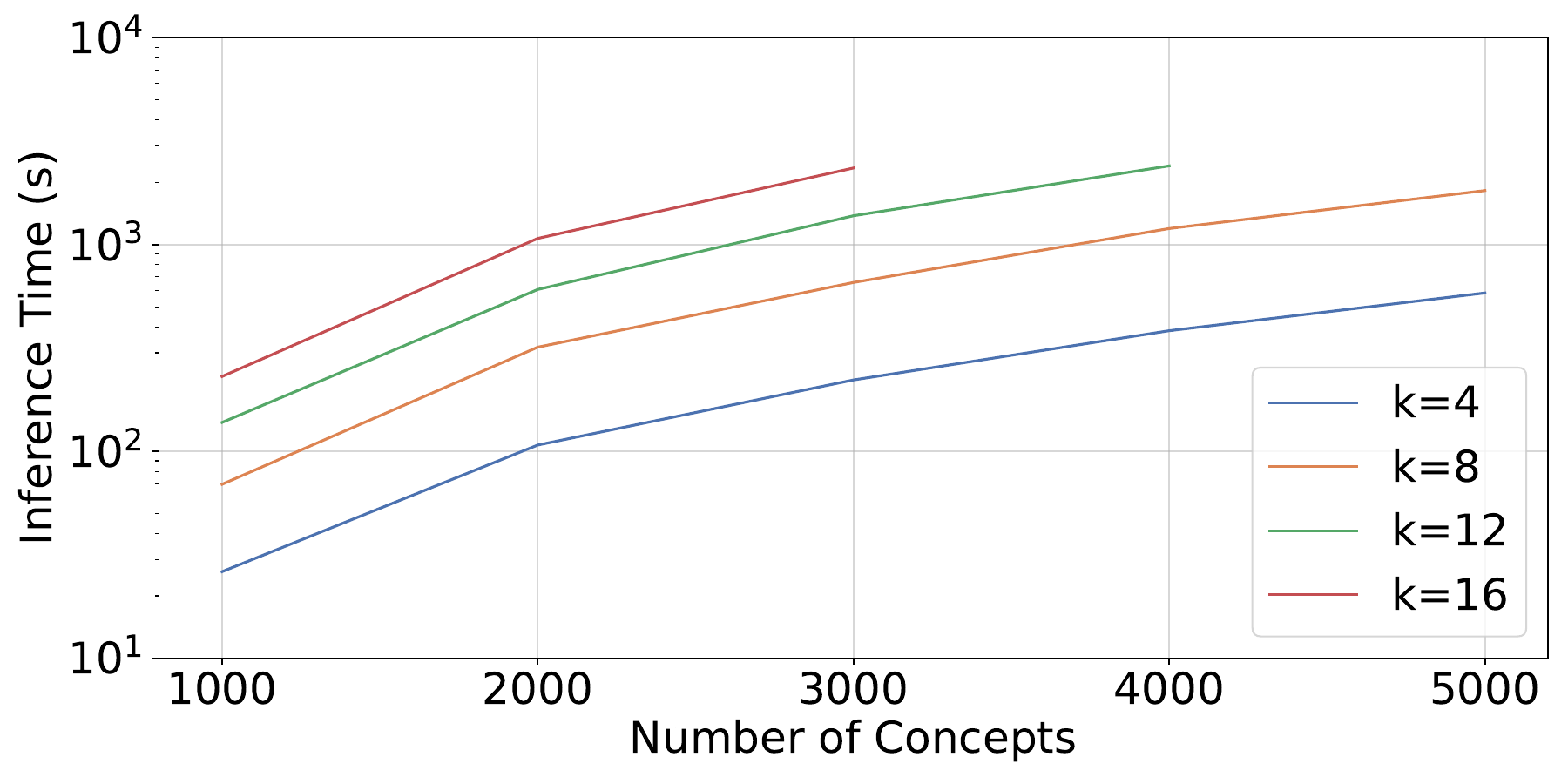}
      \caption{Scalability of MRF inference with respect to the number of concepts and the connectivity $k$. $k=12$ scales up to 4000 concepts (nearly 8 millions of random variables) and $k=16$ scales up to 3000 concepts (about 4.5 millions of random variables).}
      \label{fig:lbp_scalability}
    \end{minipage}
  \end{figure*}

\subsection{Efficiency of MRF Modeling}\label{subsec:exp_efficiency}
  \textbf{RQ3: How efficient are MRF construction and inference?}

  Figure~\ref{fig:efficiency} shows the MRF construction and inference time of different inference algorithms (i.e., pgmpy-MPLP, PGMax-CPU-LBP, and PGMax-GPU-LBP) on three datasets. All runtime numbers are reported using either a single CPU or GPU. We exclude the Shafer-Shenoy algorithm (implemented in pyAgrum) and Gibbs sampling (implemented in pgmpy) from the comparison as they do not complete execution within an hour on our smallest dataset. 
  
  The three algorithms under consideration exhibit efficient factor graph construction, with completion time of a few seconds across all datasets. The primary distinction between these algorithms lie in their inference time. GPU-accelerated LBP (i.e., PGMax-GPU-LBP) is overall the most efficient, achieving more than 2x speedup compared to CPU-based LBP (i.e., PGMax-CPU-LBP) and over two orders of magnitude speedup compared to pgmpy-MPLP on CPU. The speedup is particularly significant on the ICPSR dataset, which has over a thousand of random variables and more than twenty thousands of factors in the constructed graph. In this case, pgmpy-MPLP does not complete running within two and a half hours. In contrast, PGMax-GPU-LBP finishes the inference in just 33.4 seconds and PGMax-CPU-LBP completes in 72.1 seconds. The superior performance of PGMax-GPU-LBP can be attributed to their flat array-based implementation of LBP in JAX~\cite{jax2018github} which leverages just-in-time compilation and parallelized array operations to optimize performance on hardware accelerators. However, for smaller graphs, such as the local MRFs in the Walmart-Amazon dataset, each of which has tens of random variables due to the dataset’s sparsity, pgmpy-MPLP appears to be more efficient while PGMax-GPU-LBP remains ~1.3x faster than PGMax-CPU-LBP.

  Although PGMax-LBP inference is slower on a single CPU compared to GPU, we can batch process independent local MRFs in the Walmart-Amazon dataset across multiple CPUs. When we use all 16 CPUs on the computing node, the runtime of PGMax-CPU-LBP can be reduced from nearly 1000 seconds to under 70 seconds.
  
\subsection{Scalability}\label{subsec:exp_scalability}
  \textbf{RQ4: How does MRF inference scale as MRF grows in size with a fixed amount of memory?}

  Given that MRF inference is often the most time- and memory-intensive component of the solution, we evaluate its scalability under fixed resources: a single GPU (NVIDIA A40, 40 GB) and 256 GB of RAM. We synthesize sparse factor graphs with varying numbers of concepts (up to 5,000, translating to over 12 million random variables) and connectivity levels $k$ (up to 16) as discussed in Section~\ref{subsec:scaling_mrf_modeling}. These factor graphs consist of binary random variables with fixed prior probabilities and ternary factors with predefined potential values. We execute the PGMax-GPU-LBP algorithm using its default hyperparameters for 200 iterations.

  Figure~\ref{fig:lbp_scalability} illustrates the relationship between inference time and the number of concepts for varying connectivity levels. As expected, the inference time grows with both the number of concepts and the connectivity k. The inference times for all completed runs remain within 40 minutes. For connectivity levels $k=4$ and $k=8$,  the algorithm completes inference on graphs with 5,000 concepts in under 30 minutes. At higher connectivity levels ($k=12$ and $k=16$), the algorithm can handle graphs with up to 4,000 and 3,000 concepts, respectively, before exceeding the memory limits. Notably, a graph with 3,000 concepts corresponds to approximately 4.5 million pairs of concepts or random variables for prediction—significantly larger than the test set size of any existing public dataset for data matching or taxonomy induction. Hence,  we consider the inference algorithm good enough for handling many use cases.

\section{Related Work}\label{sec:related_work}
  \sysname proposes an MRF-based formulation for the discovery and resolution of two relationships, equivalence and parent-child, between metadata elements in data repositories. With respect to the scope of relationships, we discuss data matching, from the data management community, and taxonomy induction, from the semantic web community, which are the most relevant to our work. Moreover, we present a summary of the role of metadata integration in dataset discovery.


  \parwoindent{Data Matching} Research efforts in generalizing matching techniques have primarily focused on solutions that generalize across domains~\cite{ShragaGR20} or tasks~\cite{DBLP:journals/pacmmod/TuFTWL0JG23}. To the best of our knowledge, Unicorn, a multi-tasking model for data matching~\cite{DBLP:journals/pacmmod/TuFTWL0JG23}, represents the SOTA in generalization across tasks.  
  It supports matching tasks for data integration, such as entity matching and schema matching, by primarily focusing on equivalence relationships. For (meta)data integration, \sysname not only goes beyond equivalence relationships to discover and resolve parent-child relationships, but also our experiments show that, despite Unicorn adopting a multi-task learning approach, it is still challenging for Unicorn to  generalize to the new matching tasks and datasets of metadata integration. Parallel to multi-task learning, we demonstrate the promising potential of transfer learning using large language models with fewer than 10 billion parameters for various tasks. With parameter-efficient fine-tuning techniques such as LoRA~\cite{DBLP:journals/corr/lora}, multiple tasks can share the same base model by incorporating lightweight adapters. These adapters are significantly smaller than the base model in size yet effectively enhance result quality across downstream tasks, making this approach highly efficient and scalable.


  \parwoindent{Taxonomy Induction} 
  The taxonomy induction literature has extensively studied the problem of extracting parent-child relationships from text documents~\cite{DBLP:conf/coling/Hearst92, DBLP:conf/acl/RollerKN18, DBLP:series/synthesis-dmk/2022Shen}. 
  Various traditional frameworks have posed the taxonomy induction problem with a probabilistic formulation~\cite{DBLP:conf/acl/SnowJN06, DBLP:conf/acl/BansalBMK14}.  
  For example, Snow et al. define the objective as finding a parent-child graph that maximizes the probability of the observed evidence~\cite{DBLP:conf/acl/SnowJN06}. Their formulation relies on two additional independence assumptions to decompose the joint probability so that they can solve the problem with a heuristic search algorithm. This inspired us to design a simpler and more intuitive joint probability formulation, i.e., finding the most probable relationship graph given the existing evidence. This alternative formulation has two advantages. First, it allows us to incorporate various powerful priors, such as LLMs. Second, it allows us to cast our problem to MRF to incorporate relationship properties, such as transitivity, to refine the results. Bansal et al. also propose an MRF formulation for taxonomy induction~\cite{DBLP:conf/acl/BansalBMK14}. \sysname not only extends this formulation with an additional relationship type, it also explores different ways of obtaining prior beliefs and learning parameters for the MRF formulation. Chain-of-Layer~\cite{KhatiwadaFSCGMR23}, SOTA in the literature, iteratively prompts a LLM to construct the taxonomy while employing a smaller language model to reduce hallucinations of the LLM. However, our experiments reveal that Chain-of-Layer struggles to construct taxonomies effectively when working with a diverse set of metadata concepts from the social science domain. The deficiencies are mainly attributed to the prompts tailored for taxonomies with simple structures (e.g., country-state hierarchies) and the smaller language model's inability to generalize effectively to entities within the social science domain.

  \parwoindent{Metadata and Dataset Discovery} Metadata has been heavily used in facilitating downstream tasks such as data discovery and data sharing. Google dataset search indexes the metadata of datasets (published by users in \texttt{schema.org} format)~\cite{BrickleyBN19}. Auctus, a domain-specific and open-source dataset search engine, supports spatial and temporal data augmentation through indexing the data summaries describing dataset contents (e.g. grid size for spatial data)~\cite{CasteloRSBCF21}. Raw metadata (e.g. column names and tags)~\cite{BogatuFP020,NargesianPBZM23}, concept annotations~\cite{DBLP:journals/pvldb/NargesianZPM18}, and concept hierarchies~\cite{KhatiwadaFSCGMR23} have been used as data summaries to be indexed for table union and join search. In addition to data cataloging, the hierarchies inferred by \sysname can further enhance the results quality of downstream data discovery tasks, particularly for domain-specific data repositories. For instance, datasets tagged with metadata of varying granularity could better align with user queries, improving retrieval accuracy. Exploring these applications represents a promising direction for future work.

\section{Conclusion}
  We introduce the metadata integration problem and propose \sysname, a data-driven solution that unifies metadata concepts for a given relationship. By combining advanced prior models with probabilistic modeling and inference using Markov Random Fields, \sysname effectively resolves relationships between metadata concepts while enforcing essential relationship properties such as transitivity. Our approach also demonstrates both efficiency and scalability across various datasets. A promising direction for future work is leveraging integrated metadata vocabularies to facilitate the discovery of siloed datasets.
  




\bibliographystyle{ACM-Reference-Format}
\bibliography{reference}

\end{document}